\documentclass[12pt]{article}

\usepackage{graphics}
\usepackage{color}
\usepackage{amssymb}
\usepackage{latexsym}

\font\manual=manfnt \def\dbend{\lower3.5pt\hbox{\manual\char127}}

\def\ie{{\it i.e.}}
\def\eg{{\it e.g.}}
\def\cf{{\it c.f.}}

\def\sst{\scriptscriptstyle}

\def\frac#1#2{{#1\over#2}}
\def\coeff#1#2{{\textstyle{#1\over #2}}}
\def\half{\frac12}
\def\hf{{\textstyle\half}}

\def\IR{{\mathbb R}}

\def\IZ{{\mathbb Z}}

\catcode`\@=11
\def\slash#1{\mathord{\mathpalette\c@ncel{#1}}}
\def\underrel#1\over#2{\mathrel{\mathop{\kern\z@#1}\limits_{#2}}}

\catcode`\@=12
\def\ket#1{|#1\rangle}

\def\vev#1{\langle#1\rangle}

\def\sinh{{\rm sinh}} 	
\def\cosh{{\rm cosh}} 	
\def\tanh{{\rm tanh}}

\def\exp{{\rm exp}}

\def\LL{{\cal L}}

\def\OO{{\cal O}}

 \def\CS{{\cal S}}
 
\def\UU{{\cal U}}

\def\ZZ{{\cal Z}}

\def\unlockat{\catcode`\@=11}
\def\lockat{\catcode`\@=12}
\unlockat
\def\newsec#1{\global\advance\secno by1\message{(\the\secno. #1)}
\global\subsecno=0\global\subsubsecno=0\eqnres@t\noindent
{\bf\the\secno. #1}
\writetoca{{\secsym} {#1}}\par\nobreak\medskip\nobreak}
\global\newcount\subsecno \global\subsecno=0
\def\subsec#1{\global\advance\subsecno
by1\message{(\secsym\the\subsecno. #1)}
\ifnum\lastpenalty>9000\else\bigbreak\fi\global\subsubsecno=0
\noindent{\it\secsym\the\subsecno. #1}
\writetoca{\string\quad {\secsym\the\subsecno.} {#1}}
\par\nobreak\medskip\nobreak}
\global\newcount\subsubsecno \global\subsubsecno=0
\def\subsubsec#1{\global\advance\subsubsecno by1
\message{(\secsym\the\subsecno.\the\subsubsecno. #1)}
\ifnum\lastpenalty>9000\else\bigbreak\fi
\noindent\quad{\secsym\the\subsecno.\the\subsubsecno.}{#1}
\writetoca{\string\qquad{\secsym\the\subsecno.\the\subsubsecno.}{#1}}
\par\nobreak\medskip\nobreak}
\def\subsubseclab#1{\DefWarn#1\xdef
#1{\noexpand\hyperref{}{subsubsection}%
{\secsym\the\subsecno.\the\subsubsecno}%
{\secsym\the\subsecno.\the\subsubsecno}}%
\writedef{#1\leftbracket#1}\wrlabeL{#1=#1}}
\lockat
\newcommand{\mthb}[1]{\mbox{\boldmath \(#1\)}}

\newcommand{\be}{\begin{equation}}
\newcommand{\ee}{\end{equation}}

\newcommand{\bbb}{\begin{eqnarray}}
\newcommand{\eee}{\end{eqnarray}}
\newcommand{\pref}[1]{(\ref{#1})}

\begin{document}
 

%
\def\qtil{\tilde Q}
\def\mm{{\mthb\mu}}
\def\mub{\mu_{\!B}^{ }}
\def\pp{{\bf p}}
\def\qq{{\bf q}}
\def\xx{{\bf x}}
\def\tim{{\hat\tau}}
\def\annulus{1}
\def\dspm{2}
\def\Lpot{3}
\def\pinch{4}
%

 
\begin{titlepage}
\rightline{EFI-03-22}
 
\rightline{hep-th/0305148}
 
\vskip 3cm
\centerline{\Large{\bf The Annular Report on Non-Critical String Theory}}
 
\vskip 2cm
\centerline{
Emil J. Martinec\footnote{\texttt{e-martinec@uchicago.edu}}}
\vskip 12pt
\centerline{\sl Enrico Fermi Inst. and Dept. of Physics}
\centerline{\sl University of Chicago}
\centerline{\sl 5640 S. Ellis Ave., Chicago, IL 60637, USA}
 
\vskip 2cm
 
\begin{abstract}
Recent results on the annulus partition function 
in Liouville field theory are applied to non-critical string theory,
both below and above the critical dimension.  
Liouville gravity coupled to $c\le 1$ matter has a dual
formulation as a matrix model.  Two well-known matrix model results 
are reproduced precisely using the worldsheet formulation:
(1) the correlation function of two macroscopic loops,
and 
(2) the leading non-perturbative effects. 
The latter identifies the eigenvalue instanton amplitudes
of the matrix approach with disk instantons of the worldsheet
approach, thus demonstrating that the matrix model is
the effective dynamics of a D-brane realization of
$d\le 1$ non-critical string theory.
In the context of string theory above the critical dimension, 
\ie~$d\ge 25$, Liouville field theory realizes 
two-dimensional de~Sitter gravity on the worldsheet.
In this case, appropriate D-brane boundary conditions
on the annulus realize the S-matrix for two-dimensional de~Sitter gravity.
\end{abstract}

\end{titlepage} 
 
\newpage
 
\setcounter{page}{1}

 
\section{\label{introsec}Introduction}

Non-critical string theory has been a thread running through
the remarkable progress achieved by string theory over the years,
beginning with Polyakov's seminal work%
~\cite{Polyakov:1981rd,Polyakov:1981re}.
The study of the Liouville theory
\be
    \CS_{L} =
        \frac1{2\pi}\int \!\! dt d\sigma\sqrt{\!-\hat g}\Bigl(
        \hf(-\hat\nabla\varphi)^2-\coeff{Q}{2}\hat R\varphi
        -{2\pi\mu}\,e^{2b\varphi}\Bigr)
\label{sliou}
\ee
introduced there was a motivation for the ground-breaking work 
of~\cite{Belavin:1984vu} on two-dimensional conformal field theory,
as well as the modern treatment of covariant string perturbation theory%
~\cite{Friedan:1982is,Alvarez:1983zi,Friedan:1986ge}.
The first progress toward a non-perturbative formulation
of string theory came with the development of matrix models%
~\cite{Douglas:1990ve,Brezin:1990rb,Gross:1990vs}
of strings in $d\le 1$ target space dimensions
(see~\cite{Ginsparg:1993is} for a comprehensive review).
These matrix models foreshadowed the development of
D-branes~\cite{Polchinski:1995mt} and the AdS/CFT
correspondence~\cite{Maldacena:1998re},
in that closed strings are represented as excitations
at large $N$ of the matrix collective field. 
Indeed, we will see below that the matrix eigenvalues
are directly related to D-branes in Liouville theory.%
\footnote{This connection appeared already in~\cite{Polchinski:1994fq}
(see also~\cite{Li:1996ed}-\cite{Nakamura:2000pi}),
and seems to be a part of folklore; however, 
there have been relatively few {\it quantitative} comparisons
of Liouville D-brane amplitudes and the matrix model.
The disk one-point functions are computed in%
~\cite{Fateev:2000ik,Teschner:2000md,Zamolodchikov:2001ah},
and the open string disk two-point function is
analyzed in~\cite{Kostov:2002uq}.}

Noncritical string theory in $d>25$ is less studied%
~\cite{Myers:1987fv}-\cite{Cooper:1991zc}.
It was shown recently in%
~\cite{DaCunha:2003fm} that supercritical
string backgrounds provide models of quantum cosmology
that rather accurately model the physics
of de~Sitter space and inflation, in a context where
quantum gravity is under rather good control.
One of the main motivations for the present work was to
undertake the quantization of these models.

In spite of all the progress the Liouville theory has inspired
over the years, an understanding of its detailed properties --
for instance, a prescription to compute correlation functions --
is of relatively recent vintage%
~\cite{Dorn:1994xn,Zamolodchikov:1996aa}%
\footnote{Although the roots of the method go back to 
the early literature, see \eg%
~\cite{Gervais:1984sa,Gervais:1986cy,Dotsenko:1984nm,Dotsenko:1985ad}.}
(for a review and further references, see~\cite{Teschner:2001rv}).
A treatment of conformal boundary conditions and
boundary correlation functions is even more recent%
~\cite{Fateev:2000ik,Teschner:2000md,Zamolodchikov:2001ah}.

Our eventual goal is to apply this recent progress 
in Liouville theory to the quantization
of two-dimensional de~Sitter gravity, and in particular 
to give a prescription for what could be called the 
$2d$ de~Sitter S-matrix~\cite{Witten:2001kn,Strominger:2001pn}.
There has been much speculation recently on what
are the proper ingredients for a quantum theory
of gravity with positive cosmological constant, 
see for example%
~\cite{%
Banks:2000fe,%
Bousso:2000nf,%
Witten:2001kn,%
Strominger:2001pn},
even to the extent of questioning whether such a theory exists%
~\cite{Dyson:2002nt,Dyson:2002pf,Goheer:2002vf}.
Below we will show that, at least in two dimensions,
there is a perfectly acceptable, conventional theory
of quantum de~Sitter gravity.

By way of preparation, we take what seems like a detour into the world
of $d\le 1$ string theory, and reproduce several results
obtained quite some time ago
using the matrix model representation, here using Liouville theory
on the annulus with boundary cosmological constant(s),
coupled to appropriate boundary states of the matter CFT.
The first reproduced result is the correlation function 
of two macroscopic loops, that is, 
worldsheet boundaries of fixed length
\be
  \ell = \oint\!ds\, e^{b\varphi}
\label{bdylen}
\ee
with appropriate Dirichlet boundary conditions on the matter CFT.
We will find in section~\ref{minmod}
precise agreement with the matrix model results of
Moore and Seiberg~\cite{Moore:1992ag} for this quantity.  

The second reproduced result is the strength of the leading
non-perturbative effects in $d<1$ string theory%
~\cite{Shenker:1990uf,David:1990ge,David:1991sk,Eynard:1993sg}.
In the matrix model representation, these come from
`eigenvalue instantons', subsidiary stationary points 
of the matrix integral with one matrix eigenvalue lying a finite
distance from the endpoint of the large N eigenvalue distribution.
We will again find complete agreement with the matrix model results
in section~\ref{infbranes}, using
a different set of Liouville boundary states introduced in%
~\cite{Zamolodchikov:2001ah}
(following early work of%
~\cite{D'Hoker:1982er,D'Hoker:1983ef,D'Hoker:1983is}).
The wavefunctions of these latter boundary states are concentrated
in the strong coupling region $\varphi\to\infty$
of $c\le1$ non-critical string theory,
and hence are the appropriate states to use to compute such
non-perturbative effects.
This result identifies the eigenvalue instantons of the
matrix model with D-instantons of the worldsheet approach,
demonstrating (in agreement with folklore) that the matrix model
was in hindsight the first instance of an AdS/CFT type of duality.

With the confidence 
inspired by these calculations, we turn in section~\ref{dssec}
to the super-critical Liouville theory appropriate to
strings in $d\ge25$ dimensions 
(always excluding the Liouville direction in the definition of $d$).
Here the classical solutions of Liouville theory
describe two-dimensional de~Sitter space -- 
Liouville gravity coupled to $c\ge25$ matter is
de~Sitter gravity plus matter.
The previous calculations turn out to be less of a detour than
one might have thought.  The conformal (Carter-Penrose) diagram
of classical global $2d$ de~Sitter space 
is a finite cylinder or annulus, and the Liouville field
takes the value $\phi=\infty$ on the conformal boundary.%
\footnote{Passage to the super-critical dimension 
involves a Wick rotation $\phi=i\varphi$.}
Thus the boundary states of%
~\cite{Zamolodchikov:2001ah}
used in the D-instanton calculation are also the
appropriate boundary states for the description of 
asymptotically de~Sitter worldsheets above the critical dimension,
and the annulus worldsheet is the appropriate topology
for spacetimes that are asymptotically de~Sitter in
both the past and the future.
One-point functions on the disk describe spacetimes
with only one asymptotic de~Sitter region, and a big bang
or big crunch in the past or future.
We give a prescription for the computation of the transition
amplitudes for $2d$ spacetimes that are asymptotically de~Sitter,
and calculate them for a large class of initial conditions.

It should perhaps also be mentioned that Liouville theory
enters critical string theory in a central way,
in the description of dynamics in the throat of NS5-branes%
~\cite{Callan:1991dj,Aharony:1998ub,Giveon:1999zm,Giveon:1999px}
and in $AdS_3$~\cite{Giveon:1998ns};
the results below have interesting implications for
these applications as well, which we hope to explore elsewhere.

\vskip .5cm
{\it Note added:}  While we were writing up our results,
the work~\cite{McGreevy:2003kb} appeared,
having overlap with the material of section~\ref{infbranes}.


\section{\label{minmod}Minimal models on the annulus}

We wish to calculate the partition function on the annulus
of Liouville field theory coupled to $c\le 1$ matter.
Our immediate goal is to reproduce 
the correlation function of two macroscopic loops
of length $\ell$, $\ell'$, and matter momentum $p$,
obtained by Moore and Seiberg~\cite{Moore:1992ag}
using the matrix integral representation of these models:%
\footnote{For $c=0$ matter, this result was derived
implicitly in~\cite{Banks:1990df} and explicitly
in~\cite{Ambjorn:1990ji}; the latter also gives general
results for $n$ loops.}
\bbb
  \vev{W(\ell,p)W(\ell',-p)}
	&=& \int_0^\infty\frac{dE}{2\pi}\;G(E,p)\,\psi_E(\ell)\psi_E(\ell')
\nonumber\\
\nonumber\\
  \psi_E(\ell) &=& \frac1\pi\sqrt{E\,\sinh(\pi E)}\, K_{iE}(M\ell)
\label{msresult}\\
\nonumber\\
  G(E,p) &=& \cases{ \frac{\pi}{\cosh(\pi E)} &\quad $c=0$ \cr
		\frac{\pi}{\cosh(\pi E)\mp\half} &\quad $c=\half$ \cr
		\frac{1}{E^2+p^2}\frac{2\pi E}{\sinh(\pi E)} 
			& \quad $c=1$ \quad.}
\nonumber
\eee
The quantity $M$ is proportional to $\sqrt\mu$;
one is free to absorb the constant into the definition of $\ell$,
but we will not do so.
For $c=0$, there are no matter degrees of freedom; the
only allowed matter momentum is $p=0$.  Similarly,
for the Ising model $c=\hf$, there are two possible 
Dirichlet boundary states of fixed boundary spin,
and hence two values of the loop momentum.
For $c=1$, the momentum $p\in \IR$ if the target is non-compact.%
\footnote{We have changed the normalization of the $c=1$ propagator
relative to the answer in~\cite{Moore:1992ag} by a factor of two,
in order to make it consistent with the $c<1$ formulae.}
It was pointed out in~\cite{Kostov:1993am} that 
if one makes the $c=1$ result periodic in momentum space
\be
  \sum_{n=-\infty}^\infty \frac{1}{E^2+(p+2n)^2}\,\frac{2 E}{\sinh(\pi E)}
	=\frac{\pi}{[\cosh(\pi E)-\cos(\pi p)]}\ ,
\label{chain}
\ee
all the amplitudes have the same basic structure
of momentum space propagator.
This form of the $c=1$ propagator
is appropriate for a target space which is
a one-dimensional lattice $\IZ$ rather than the continuous line $\IR$.

In the worldsheet represtentation of string perturbation theory,
the amplitudes~\pref{msresult} arise from the annulus topology.
In the standard conformal gauge,
the annulus partition function for $2d$ gravity 
coupled to matter is of the form 
\be
  \ZZ = \int_0^\infty \!d\tau\; \ZZ_{\rm ghost}\;\ZZ_{\rm Liouville}
	\;\ZZ_{\rm matter}\quad.
\label{genzann}
\ee
Here $\tau$ is the annulus modulus in the {\it closed} string channel
(see figure~\annulus).

\begin{figure}[ht]
\begin{center}
\[
\mbox{\begin{picture}(129,84)(0,0)
\includegraphics{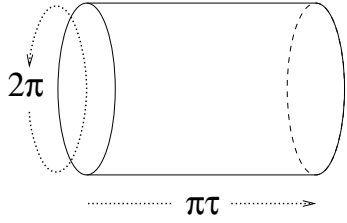}
\end{picture}}
\]
\caption{\it 
Finite cylinder worldsheet for closed string propagation.
}
\end{center}
\end{figure}

The various constituent partition functions are known.
First, the partition function for the Faddeev-Popov
ghosts of conformal gauge is%
~\cite{Polchinski:1998rq}
\be
  \ZZ_{\rm ghost}=\eta(\qq)^2\ , 
\label{zghost}
\ee
where $\eta(\qq)$ is the usual Dedekind eta function,
$\eta(\qq)=\qq^{1/24}\prod_n(1-\qq^n)$,
and as usual $\qq=\exp[-2\pi \tau]$.

In conformal gauge, $2d$ gravity dynamics is that of
the scale factor of the metric, and 
is governed by the Euclidean Liouville action 
\be
  \LL = \frac{1}{4\pi} \int\!d^2\!z\;\Bigl(
	(\partial\varphi)^2+QR\,\varphi +4\pi\mu e^{2b\varphi}\Bigr)
	+\oint \!ds\Bigl(\frac{1}{2\pi}QK\,\varphi + \mub e^{b\varphi}\Bigr)\ ,
\label{liouact}
\ee
where we have included the boundary interactions
with the extrinsic curvature $K$ as well as a boundary
cosmological constant $\mub$ (the latter can be different
on each disconnected boundary component).
The Liouville partition function 
was determined up to normalization in recent work 
of Fateev, the Zamolodchikovs, and Teschner%
~\cite{Fateev:2000ik,Teschner:2000md,Zamolodchikov:2001ah}
\bbb
  \ZZ_{\rm Liouville} &=&
	\int \!d\nu\; \Psi_{\nu}(\sigma)\Psi_{-\nu}(\sigma')\chi_\nu(\qq)
\label{zliou}\\
  \Psi_\nu(\sigma) &=&
	\frac{\Gamma(1+2i\nu b)\Gamma(1+2i\nu /b)
	\cos(2\pi\sigma \nu )}%
		{2^{1/4}\;(-2i\pi \nu) }\;\mm^{-i\nu /b}\ .
\label{Lfunction}
\eee
The notation is as follows:
the parameters defining the theory are characterized by
\bbb
  c_L &=& 1+6Q^2\quad ,\qquad 
  Q=b^{-1}+b\quad ;
\nonumber\\
  \mm &=& \pi\mu\gamma(b^2)\ ;
\label{cdef}\\
  \mub &=& \frac{\sqrt\mu\;\cosh(\pi b\sigma)}{\sqrt{\sin(\pi b^2)}}
	= \frac1\pi\Gamma(1\!-\!b^2)\sqrt\mm\,\cosh(\pi b\sigma)\ ;\quad
\nonumber
\eee
with $\gamma(x)=\Gamma(x)/\Gamma(1-x)$.
The quantity $\mm$
is the (KPZ) scaling parameter of the correlation functions.
Finally, the character of a non-degenerate Liouville primary is
\be
  \chi_\nu (\qq) = \frac{\qq^{\nu^2}}{\eta(\qq)}\ .
\label{liouchar}
\ee
We consider first $c=1$ matter, then the slightly
more complicated case of $c<1$.


\subsection{\label{conesubsec}c=1}

For $c=1$ matter on a continuous, non-compact target, the matter
partition function is
\be
  \ZZ_{\rm matter} =
  \int_{-\infty}^\infty\! {dp}\, 
	\frac{1}{\sqrt2}\;\frac{\qq^{p^2/4}}{\eta(\qq)}
	\, e^{ip(x-x')}
	\ .
\label{conematter}
\ee
The factor of $1/\sqrt2$ is the square of the Dirichlet
boundary entropy~\cite{Elitzur:1998va}.%
\footnote{The analysis of~\cite{Elitzur:1998va}
employs the convention $\alpha'=\half$; in
the present work we are setting $\alpha'=1$.}
Combining the component partition functions, the factors
of $\eta(\qq)$ cancel.  The integral over $\tau$ yields
\be
  \ZZ(\sigma,x|\sigma',x') = 
	\int\!\frac{dp}{2\pi}\int\! d\nu\, 2\;
	\frac{e^{ipx}\cos(2\pi\sigma\nu)\cdot e^{-ipx'}\cos(2\pi\sigma'\nu)}%
		{\sinh^2(2\pi\nu)[(2\nu)^2+p^2]}
\label{conereal}
\ee
where we have set $b=1$ for $c_{\rm matt}=1$.

To compare to the result~\pref{msresult}, we note that
\bbb
  K_{2i\nu/b}(M\ell) 
	&=& \int_0^\infty \!d(\pi b\sigma)\; e^{-M\ell\,\cosh(\pi b\sigma)}
		\,\cos(2\pi \nu\sigma)
\nonumber\\
  \frac{\pi\,\cos(2\pi \nu\sigma)}{(2\nu/b)\,\sinh(2\pi \nu/b)}
	&=& \int_0^\infty\!\frac{d\ell}{\ell}\,
		e^{-M\ell\,\cosh(\pi b\sigma)}\,
		K_{2i\nu/b}(M \ell)\quad .
\label{backlund}
\eee
Note that, since the boundary interaction is
$\oint\!\mu_{\!B}^{ }\, e^{b\varphi}\!=\!\mu_{\!B}^{ }\ell$,
we can now identify the parameter $M$ 
in~\pref{msresult} as $M\!=\!\pi^{-1}\Gamma(1\!-\!b^2)\sqrt\mm$,
which reduces to $\sqrt\mm/\pi$ in the semi-classical limit $b\to0$.

The correlator~\pref{msresult} can be Laplace transformed
to a function of the boundary cosmological constants
$\mub= M\cosh(\pi b\sigma)$
(see equation~\pref{cdef})
using the second relation of~\pref{backlund}; the partition
function in the worldsheet approach is most
naturally given as a function of $\mub$.
The two-loop correlator~\pref{msresult} becomes%
\footnote{The correlator as a function of loop length is
in fact better defined, and regulates the divergence
at small $E$ in equation~\pref{conereal}; a better comparison
would therefore take the inverse transform of~\pref{conereal}
and compare it directly to~\pref{msresult}.}
\be
  \langle{W(\sigma,p)W(\sigma',-p)}\rangle_{c=1} =
	\int_0^\infty\! {dE}\;
	\frac{\cos(\pi \sigma E)\cdot\cos(\pi \sigma' E)}%
		{\sinh^2(\pi E)\,[E^2+p^2]}\quad.
\label{mstransfcone}
\ee
Identifying $E=2\nu$, 
this is exactly the result~\pref{conereal}.

\subsection{\label{clessone}c$<$1}

Let us now turn to $c<1$ minimal models coupled to gravity.
The annulus partition function of $c<1$ matter is known from
work of Saleur and Bauer~\cite{Saleur:1989zx}.
The $c\le1$ models are parametrized by
\be
  c_m=1-6\qtil^2\quad,\qquad Q=\tilde b^{-1}-\tilde b\quad,\qquad
	\tilde b=\sqrt{q/p} 
\label{matparam}
\ee
for $p,q\in\IZ$.  Coupling to gravity sets $b=\tilde b$.  
The unitary matter models have \mbox{$p=q\pm1$} and
can be thought of as the conformal continuum limit of 
a lattice height model, where the heights are associated
to the nodes of an $ADE$ Dynkin diagram~\cite{Pasquier:1987jc}.
In the context of coupling to quantum gravity,
these models were studied 
using matrix model techniques, in a series of works by 
Kostov and collaborators%
~\cite{Kostov:1989eg}-\cite{Kostov:1996xw},~\cite{Kostov:1993am},%
~\cite{Kostov:2002uq}.
The basic idea is to put a gas of loops on a 
fluctuating random lattice; the loops are the domain boundaries
between different heights on the Dynkin diagram. 
In~\cite{Kostov:2002uq} (see also~\cite{Kazakov:1992pt}), 
some of the boundary correlators are worked out, 
and the phase of the loop gas model that
connects to the continuum path integral is identified --
the Liouville theory with action~\pref{liouact}
is described by a dilute gas of loops,
and the boundary interaction
in~\pref{liouact} is appropriate to a Dirichlet boundary
condition on the heights.  The dilute loop gas corresponds
to the condition $q<p$.  The matter annulus partition function
in this phase, with fixed heights $a,c$ on the two boundaries, is given
by~\cite{Saleur:1989zx} 
\bbb
  \ZZ_m(a|c) &=& \sum_{r=1}^{q-1}\sum_{j\in{\it exp}}
	\,S_c^{(j)}S_a^{(j)}\,
	\frac{\sin(\frac{\pi rj}{q})}{\sin(\frac{\pi j}{q})}\,
		\chi_{r,1}(\qq')
\nonumber\\
  \chi_{r,s}(\qq') &=&  
	\frac{1}{\eta(\qq')}
	\sum_{k=-\infty}^{\infty} [(\qq')^{\alpha_{r,s}(k)}
		-(\qq')^{\alpha_{r,-s}(k)}]
\label{matz}\\
  \alpha_{r,s}(k) &=& \frac{(2pqk+pr-qs)^2}{4pq}
\nonumber
\eee
where $j$ runs over the exponents of the Dynkin diagram,
which we take to be the $A_{q-1}$ diagram.
The matter wavefunctions
\be
  S_a^{(j)}=\sqrt{2/q}\,\sin(\pi ja/q)
\label{Dynkfn}
\ee
are eigenfunctions of the $A_{q-1}$ Cartan matrix,
which is the discrete Laplace operator on the Dynkin diagram.

Now, the above partition function is written in the 
open string channel $\qq'=\exp[-2\pi/\tau]$;
we need to rewrite it in the closed string channel
in which the Liouville and ghost partition functions~\pref{zliou}
are written.  For this purpose, we may use a formula
from~\cite{Cardy:1986ie}
\be
  \chi_{r,s}(\qq') = \frac{1}{\eta(\qq)}
		\Bigl(\frac{2}{pq}\Bigr)^{1/2}
	\sum_{l=-\infty}^\infty \exp\Bigl[\frac{-2\pi\tau\,l^2}{4pq}\Bigr]
		\sin\Bigl(\frac{\pi rl}{q}\Bigr)\,
		\sin\Bigl(\frac{\pi sl}{p}\Bigr)\ ;
\label{modtransf}
\ee
writing $l=2kq+k'$, the sum over $r$ gives
\be
  \sum_{r=1}^{q-1} \sin\Bigl(\frac{\pi rl}{q}\Bigr)
		\,\sin\Bigl(\frac{\pi rj}{q}\Bigr)
	= \frac q2\,(\delta_{k',j}-\delta_{k',-j})
\label{spinsum}
\ee
and so
\be
 \ZZ_m(a|c) = \sum_{j=1}^{q-1}
        \frac{ S_c^{(j)}S_a^{(j)} }{\sin(\frac{\pi j}{q})}\,
	\Bigl(\frac{q}{2p}\Bigr)^{\half} 
	\sum_{k=-\infty}^\infty\left[
	\qq^{(2kq+j)^2/4pq}\,\sin[\coeff{(2kq+j)\pi}{p}]
	-\qq^{(2kq-j)^2/4pq}\,\sin[\coeff{(2kq-j)\pi}{p}]\right]\ .
\label{zmatclosed}
\ee

We are now ready to combine the component partition functions
and integrate over $\tau$.  The factors of $\eta(\qq)$
cancel, and one is left with simply the exponential factors~--
the gaussian sum of powers of $\qq$ in~\pref{zmatclosed},
and a factor $\qq^{\nu^2}$ from~\pref{liouchar}.
The result of the $\tau$ integral is
\bbb
  \ZZ(a|c) &=& \sum_{j=1}^{q-1}\int_0^\infty\!\!{d\nu}\;
	\frac{\cos(2\pi\sigma \nu)\cos(2\pi\sigma'\nu)}%
		{\sqrt2\;\sinh(2\pi\nu/b)\,\sinh(2\pi\nu b)}
\label{zint}\\
  & &\hskip .3cm
        \times\;\frac{ S_c^{(j)}S_a^{(j)} }{\sin(\frac{\pi j}{q})}\,
        \Bigl(\frac{q}{2p}\Bigr)^{\half}
        \frac{2}{2\pi}\left\{
		\sum_{k=1}^\infty\left[
		\frac{\sin[\coeff{(2kq+j)\pi}{p}]}{\frac{(2k+j)^2}{4pq}+\nu^2}
		-(j\to -j)\right]+
		\frac{\sin[\coeff{j\pi}{p}]}{\frac{j^2}{4pq}+\nu^2}\right\}\ .
\nonumber
\eee
One is now instructed to expand out the argument of the sine
in the numerator of the sum over $k$, and also to expand the denominators
into a sum over simple poles and recombine terms.  
The expression in curly brackets in~\pref{zint} becomes
\bbb
  && \sum_{k=1}^\infty \frac{\cos[\frac{j\pi}{p}]}{2i\nu\cdot 2\beta q}\left[
	\frac{2k\sin[\frac{2kq\pi}{p}]}%
		{k^2+(\frac{\nu}{2\beta q}+i\frac{j}{2q})^2}
	-\frac{2k\sin[\frac{2kq\pi}{p}]}%
                {k^2+(\frac{\nu}{2\beta q}-i\frac{j}{2q})^2}\right]
\label{intermediatesum}\\
&&\nonumber\\
  &&\hskip 2cm + \sum_{k=1}^\infty 
	\frac{\sin[\frac{j\pi}{p}]}{2i\nu\cdot 2\beta q}\left[
	\frac{\frac{2(-\beta j+i\nu)}{2\beta q}\,\cos[\frac{2kq\pi}{p}]}%
	{k^2+(\frac{\nu}{2\beta q}+i\frac{j}{2q})^2} +
	\frac{\frac{2(\beta j+i\nu)}{2\beta q}\,\cos[\frac{2kq\pi}{p}]}%
        {k^2+(\frac{\nu}{2\beta q}-i\frac{j}{2q})^2}\right]
  +~\frac{\sin[\coeff{j\pi}{p}]}{\beta^2j^2+\nu^2}
\nonumber
\eee
where we have defined $\beta^2=(4pq)^{-1}$.
Now we make use of the identities
\bbb
  \sum_{k=1}^\infty \frac{k\sin(kx)}{k^2+\alpha^2}
	&=& \frac{\pi}{2}\;\frac{\sinh[\alpha(\pi-x)]}{\sinh[\alpha\pi]}
\nonumber\\
  \sum_{k=1}^\infty \frac{\cos(kx)}{k^2+\alpha^2}
        &=& \frac{\pi}{2\alpha}\;\frac{\cosh[\alpha(\pi-x)]}{\sinh[\alpha\pi]}
		-\frac{1}{2\alpha^2}\ ;
\label{idents}
\eee
after some algebra involving repeated use of trigonometric
identities,
one finds that~\pref{intermediatesum} reduces to
\be
  \frac{\pi}{b\,\nu}\;\frac{\sinh(2\pi \nu/b)\,\sin(\frac{j\pi}{q})}%
	{[\cosh(2\pi \nu/b)-\cos(\frac{j\pi}{q})]}
\label{simpler}
\ee
(recall $b=\sqrt{q/p}$).
Replacing the curly brackets in~\pref{zint} by this expression, we have
\be
  \ZZ(a,\sigma|c,\sigma') =
	\sum_{j=1}^{q-1}\int\!\frac{d\nu}{b}\, 
	\frac{ S_c^{(j)}\cos[2\pi\sigma \nu]
		\cdot S_a^{(j)}\cos[2\pi\sigma' \nu]}
	{(2\nu/b)\,\sinh(2\pi \nu/b)\cdot
	[\cosh(2\pi \nu/b)-\cos(\pi j/q)]}	\quad .
\label{altogether}
\ee

Now, this is indeed the right answer!  
The correlator~\pref{msresult} should again be Laplace transformed
to a function of the boundary cosmological constants
$\mub=\frac{\sqrt{\mu}\;\cosh(\pi b\sigma)}{\sqrt{\sin(\pi b^2)}}$
(see equation~\pref{cdef});
the correlator becomes
\be
  \vev{W(\sigma,p)W(\sigma',-p)} =
	\int_0^\infty\! \frac{dE}{2\pi}\,
	\frac{\pi\,\cos(2\pi b\sigma E)\cdot\cos(2\pi b\sigma' E)}%
		{E\,\sinh(\pi E)\,[\cosh(\pi E)-\cos(\pi p)]}\quad.
\label{mstransf}
\ee
Identifying $E=2\nu/b$, $p=j/q$, 
this is exactly the result~\pref{altogether}.
In particular, pure $2d$ gravity is the minimal model $p=3$, $q=2$
coupled to gravity.  Thus $\cos[\pi j/q]=0$
and we recover the result~\pref{msresult}.
Similarly, for the Ising model $p=4$, $q=3$
one has $\cos[\pi j/q]=\pm\hf$ for $j=1,2$;
again one recovers the result~\pref{msresult}.

 
\section{\label{infbranes}Branes at infinity}

Thus, matrix model results are fully compatible with
the Liouville bootstrap of%
~\cite{Fateev:2000ik,Teschner:2000md,Zamolodchikov:2001ah}.
The generic, non-degenerate Virasoro representation
of the Liouville theory yields the character~\pref{liouchar}.
One then reconstructs the annulus partition function~\pref{zliou}
from the wavefunction~\pref{Lfunction}; the latter corresponds
to the boundary state
\be
  \ket{B_\sigma} = \int_{-\infty}^\infty\!d\nu\,
	e^{2\pi i\nu\sigma}\Psi_\nu(\sigma)\;\ket{\nu}\ ,
\label{lbdystate}
\ee
where $\ket{\nu}$ are Ishibashi states%
~\cite{Fateev:2000ik,Teschner:2000md,Zamolodchikov:2001ah,%
Ishibashi:1989kg,Onogi:1989qk}.
In the semiclassical limit $b\to 0$, these states implement the
natural boundary condition
\be
  \partial_n\varphi=QK+2\pi\mub\,b\, e^{b\varphi}
\label{lbc}
\ee
derived from the variational principal on~\pref{liouact}.
The work of~\cite{Zamolodchikov:2001ah}
also introduced another set of boundary states
(for early work, see~\cite{D'Hoker:1982er,D'Hoker:1983ef,D'Hoker:1983is})
\bbb
  \ket{m,n} &=& \int \!d\nu\, \Psi_\nu(m,n)\;\ket{\nu}
\label{mnvac}\\
  \Psi_\nu(m,n) &=& 
	2\,\sinh(2\pi m\nu/b)\,\sinh(2\pi n\nu b)\,
	\left[\frac{\Gamma(1+2i\nu b)\Gamma(1+2i\nu /b)}%
	{2^{1/4}\,(-2i\pi \nu)}\;\mm^{-i\nu /b}\right]
\nonumber
\eee
associated to degenerate representations of conformal dimension
\be
  h_{m,n} = \coeff14 Q^2 -\coeff14(m/b+nb)^2
\label{condim}
\ee
having a null vector on level $mn$.
The (normalized) disk one point function of Liouville exponentials 
with this boundary state is
\bbb
  \vev{e^{2\alpha\varphi}} &=& \frac{U_{m,n}(\alpha)}{(1-z\bar z)^{2h_\alpha}}
\nonumber\\
  U_{m,n}(\alpha) &=& \frac{\Psi_{i(Q/2-\alpha)}(m,n)}{\Psi_{iQ/2}(m,n)}
\label{onepoint}
\eee
where $h_\alpha=\coeff14 Q^2+\nu^2$.  
In particular, one has
\be
  \vev{e^{2b\varphi}} = \frac{Q}{\pi\mu b}\,\frac{1}{(1-z\bar z)^2}
\label{Lmetric}
\ee
which is the constant negative curvature metric on the Poincar\'e disk.
The boundary state~\pref{mnvac} is thus concentrated
in the region of $\varphi\to\infty$.
The amplitudes~\pref{onepoint}
have been compared~\cite{Zamolodchikov:2001ah}
with the perturbative expansion of 
Liouville theory in the semiclassical limit $b\to 0$;
agreement was found for $m=n=1$ but not otherwise.
Thus, in Euclidean Liouville theory,
the $m=n=1$ boundary state satisfies all the necessary criteria
to be identified with the boundary state at
conformal infinity of Euclidean $AdS_2$.
In particular, its null vector at level $mn=1$ is associated
to invariance under $L_{-1}$; it is the natural $SL(2,\IR)$
invariant boundary condition for Euclidean $AdS_2$.
The one-point functions~\pref{onepoint} for $m>1$ are singular
in this limit, indicating that there is no associated
semi-classical geometry for these states.
The geometrical interpretation of the states $m=1$, $n>1$
has until now also remained unclear; 
in analogy to D-branes in current algebra models,
one expects them to be some sort of bound states 
of the basic $m\!=\!n\!=\!1$ branes.

An important feature of the boundary states~\pref{mnvac}
is the spectrum of open string vertex operators that
couple to them.  It was demonstrated in~\cite{Zamolodchikov:2001ah}
that the transform to the open string channel 
of the annulus partition function
\be
  \ZZ_L(m,n|m',n') = \int\!d\nu\,\chi_\nu(\qq)
	\Psi_\nu(m,n)\,\Psi_{-\nu}(m',n')
\label{mnmn}
\ee
contains only open string characters of the
degenerate Virasoro representations $(m'',n'')$ that appear in the
fusion algebra of the degenerate representations $(m,n)$ and $(m',n')$
\be
  m'' \in \{|m'-m|+1,\dots,m+m'-1\} \quad,\qquad
  n'' \in \{|n'-n|+1,\dots,n+n'-1\} \quad.
\label{fusion}
\ee
There are thus only a finite number of 
Virasoro primaries flowing in the Liouville open string channel
with these boundary conditions.

The boundary states~\pref{mnvac} are formally related to the `standard'
boundary state~\pref{lbdystate} as
\bbb
  \ket{m,n} &=& \ket{B_{\sigma(m,n)}} -\ket{B_{\sigma(m,-n)}} 
\label{bsrel}\\
  \sigma(m,n) &=& i\Bigl(\frac mb + nb\Bigr)\ .
\nonumber
\eee
Using~\pref{cdef}, we have
\be
  \mub(m,n) = (-1)^m\sqrt\mu\;\frac{\cos(\pi n b^2)}{\sqrt{\sin(\pi b^2)}}\ ;
\label{mubmn}
\ee
note that the two values $\sigma(m,\pm n)$ contributing to~\pref{bsrel}
yield the same value of $\mub$, even though the
corresponding boundary states are distinct.

In the matrix model formulation of%
~\cite{Kostov:1989eg}-%
\cite{Kostov:1996xw},%
\cite{Kostov:2002uq}
of strings propagating along a Dynkin diagram,
the boundary cosmological constant
parameter $\mub$ of Liouville theory can be identified
with the coordinate $z$ of the complex eigenvalue plane of the matrices%
~\cite{%
Martinec:1991ht,%
Moore:1992ag,%
Ginsparg:1993is}.
In the matrix model, strings are excitations of
the large $N$ collective field of the matrices. 
The string background corresponds to a matrix eigenvalue distribution
supported along the real axis, on the interval 
\be
  z \in
	(-\infty,
	-\mub(\sigma\!=\!0))= (-\infty,-M)\quad .
\label{eigenvalcut}
\ee
The disk partition function and all disk correlation functions 
are meromorphic functions of $z$ 
with a cut along the real axis extending along
the interval~\pref{eigenvalcut}.
For instance, consider the the amplitude for a macroscopic loop 
(disk partition function at fixed boundary length $\ell$)
located at a fixed node of the Dynkin diagram of
the unitary height model; it has the factorized form%
~\cite{Kostov:1992cg}
\be
  \vev{W(a,\ell)} = w(\ell)\,\sin(\coeff{\pi a}{q})\ ,
\label{loopvev}
\ee
where the Laplace transform of the loop amplitude $w(\ell)$ yields
\bbb
  \hat w(z) &=& \int_0^\infty\!d\ell\, e^{-z\ell} w(\ell)
\nonumber\\
	&=& {\rm const.}\times
	[(z+\sqrt{z^2-M^2})^g + (z-\sqrt{z^2-M^2})^g]
\label{macroloop}
\eee
with $g=b^{-2}$.  
The quantity $\hat w(z)$ indeed has a cut on the 
physical sheet along the interval~\pref{eigenvalcut},
and the discontinuity along the cut gives the density of eigenvalues
in the matrix model~\cite{Kostov:1991hn}.
The cut is nicely parametrized by~\pref{cdef},
\ie\ $\mub\to\sigma$ is a uniformizing map that
smooths the branch point.  In terms of $\sigma$,
\be
  \hat w(\sigma) = {\rm const.}\times \cosh(\pi\sigma/b)
\label{wsig}
\ee
is analytic in $\sigma$.

The pair of boundary states in~\pref{bsrel}
live at points near the endpoint of the cut 
in the eigenvalue distribution,
on the $m^{\rm th}$ sheet.%
\footnote{The authors of~\cite{Zamolodchikov:2001ah}
were unable to associate a semiclassical, geometrical
interpretation to the states with $m>1$;
perhaps this is related to the fact that they
are not on the first, physical sheet of the $z$ plane.}
The singularity at $z=-\mub(\sigma\!=\!0)$
is associated to surfaces of diverging boundary length,
and thus $\varphi\to\infty$ near the boundary.

We can now give another interpretation to the $m,n$ boundary
states -- they are associated to D-instantons in 
$c\le 1$ non-critical string theory,  
and the eigenvalues of the matrix model are 
essentially the D-branes of Liouville theory
in a non-critical gauge/gravity correspondence.
One might regard the $m,n$ boundary states~\pref{bsrel}
as D-instantons of Liouville theory located in the 
region of large positive $\varphi$ (semiclassically, $\varphi\to\infty$),
corresponding to stationary matrix eigenvalues located outside
the cut~\pref{eigenvalcut}.

The leading non-perturbative effects in $c\le 1$ non-critical
string theory come precisely from eigenvalue instanton amplitudes
in the matrix model representation%
~\cite{Shenker:1990uf,David:1990ge,David:1991sk}, see also%
~\cite{Eynard:1993sg,Fukuma:1997hj,Fukuma:1999tj}.
In the continuum worldsheet approach, the D-instanton amplitude
is a disk partition function with appropriate boundary conditions.
We can obtain this amplitude from the annulus partition function
by factorization on the identity in the closed string channel;
this gives the absolute square of the disk instanton amplitude.
The matter partition function of interest is~\pref{matz}.
We wish to factorize it on the (gravitationally dressed)
identity operator contribution in the closed string channel; 
to do this, 
let us couple the matter partition function~\pref{zmatclosed} 
to the ghost partition function~\pref{zghost} and the Liouville
partition function with boundary conditions
corresponding to the $(m,n)$-type states~\pref{mnvac} 
on both boundaries of the annulus%
~\cite{Zamolodchikov:2001ah}
\bbb
  \ZZ_L(mn|m'n') &=& 
\label{lann}\\
  & &\hskip -2.5cm
\int_{-\infty}^\infty\!{d\nu}\, \chi_\nu(\qq)\,
	\frac{2\,\sinh(2\pi m\nu/b)\sinh(2\pi n\nu b)\cdot
		2\,\sinh(2\pi m'\nu/b)\sinh(2\pi n'\nu b)}
	{\sqrt2\;\sinh(2\pi\nu/b)\sinh(2\pi\nu b)}\ .
\nonumber
\eee
The integral over $\qq$ and resulting sum over descendants
proceeds as before, yielding
\bbb
  \ZZ(a,mn|c,m'n') &=&
\label{altogetheragain}\\
  & &\hskip -3cm
	\sum_{j=1}^{q-1}\int_{-\infty}^\infty \!\frac{d\nu}{b}\, 
	\frac{2S_c^{(j)}\sinh(2\pi m\nu/b)\sinh(2\pi n\nu b)
		\cdot 2S_a^{(j)}\sinh(2\pi m' \nu/b)\sinh(2\pi n'\nu b)}
	{(2\nu/b)\,\sinh(2\pi \nu/b)\cdot
	[\cosh(2\pi \nu/b)-\cos(\pi j/q)]}	\quad .
\nonumber
\eee
To pick out the contribution of the identity operator, 
deform the contour of integration
to pick up the poles on the positive imaginary axis.
The leading pole at $\nu = {ij}/{\sqrt{4pq}}$
is the contribution of the $j^{\rm th}$ order parameter
in the closed string channel, with $j=1$ corresponding to 
the (gravitationally dressed) lowest dimension operator,
and $j=p-q$ for the (gravitationally dressed) identity operator;%
\footnote{More generally, $j=|pr-qs|$ for the contribution
of the $(r,s)$ matter primary.}
the higher poles at fixed $j$ give the contributions of 
`gravitational descendants' of the $j^{\rm th}$ order parameter
in the terminology of matrix models.
Setting $j=p-q$ for the contribution of the identity operator, 
and furthermore putting $m=m'$, $n=n'$, and $a=c$,
the residue of the leading pole
gives the square of the disk one-point function
of the cosmological constant operator (the `puncture operator'
in matrix model terminology) with the given boundary conditions.
One finds (again recalling $b=\sqrt{q/p}$)
\be
  \ZZ_{\rm disk}(a;m,n) = \frac{\partial\Gamma_{\rm inst}}{\partial\mm} =
	\Bigl[2\sqrt2\;\frac{\sin(\frac{\pi a}{q})}{\sin(\frac{\pi j}{q})}\;
	\sin(\frac{\pi jm}{q})\,\sin(\frac{\pi jn}{p})\Bigr]
		\;\mm^{\frac{j}{2q}}
\label{instamp}
\ee
up to an overall phase.  We have also restored 
the power of $\mm$ that drops out of the $I_\nu K_\nu$
expansion of the two-loop correlator%
~\cite{Moore:1991ir,Moore:1992ag},
as one sees from the wavefunction~\pref{mnvac}.

This result should be compared with
the instanton contributions to the matrix model
evaluated in%
~\cite{Eynard:1993sg,Fukuma:1997hj,Fukuma:1999tj}:
\be
  \Gamma_{\rm inst} = 
	\Bigl[2\sqrt2\,\sin(\frac{\pi k}{q})\,\sin(\frac{\pi l}{p})\Bigr]
	\;\frac{2q}{2q+1} \,\mm^{\frac{2q+1}{2q}}
\label{nonpert}
\ee
(we have divided the result quoted in section 5.3 of%
~\cite{Eynard:1993sg} by a factor $\sqrt2$
to restore the conventional normalization
of the matrix model susceptibilities 
$u_j=\frac{\partial}{\partial\mm}\vev{\OO_j}$).
The result~\pref{nonpert} holds for unitary matter models 
$p=q+1$ coupled to gravity.
The comparison for non-unitary minimal models is
complicated by the fact that the identity operator is 
not the leading contribution to the factorization of
the annulus amplitude -- there are operators of
negative dimension that contribute stronger singularities
in the matter partition sum, and so one must be careful
in the identification of the contribution of the puncture
operator.  For simplicity, let us therefore in the comparison restrict
consideration to the unitary models $p=q+1$ (so that $j=p-q=1$)
in~\pref{instamp}.
One finds agreement with~\pref{nonpert}
after integrating~\pref{instamp} with respect to the
cosmological constant, and setting $m=1$.

To summarize, we have succeeded in reproducing a number
of basic properties of non-critical string theory,
originally derived in the context of matrix models,
using the continuum worldsheet approach.
This reinforces our understanding of how these two approaches
are dual representations of the same theory.
In particular, we have obtained a precise identification
of single eigenvalues of the matrix model with D-branes
of the worldsheet approach, demonstrating that the 
matrix models of non-critical string theory are 
in fact instances of the AdS/CFT correspondence.
Further investigation might allow
a rather detailed duality map to be developed,
which would certainly elucidate the general matrix/string duality.
Instead, we now proceed to analytically continue the result~\pref{lann}
for the Liouville annulus amplitude, in order to obtain 
some new results -- namely, a prescription for
the dynamics of two-dimensional de~Sitter quantum gravity.


\section{\label{dssec}The de~Sitter regime}

We now turn to string theory above the critical dimension,
which was shown in%
~\cite{DaCunha:2003fm}
to rather accurately model the cosmology of de~Sitter
space and inflation.  The annulus amplitude in this case
realizes the conformal (Carter-Penrose) diagram of certain regimes
of the cosmology, see for example figure~\dspm.

\begin{figure}[ht]
\begin{center}
\[
\mbox{\begin{picture}(267,167)(0,0)
\includegraphics{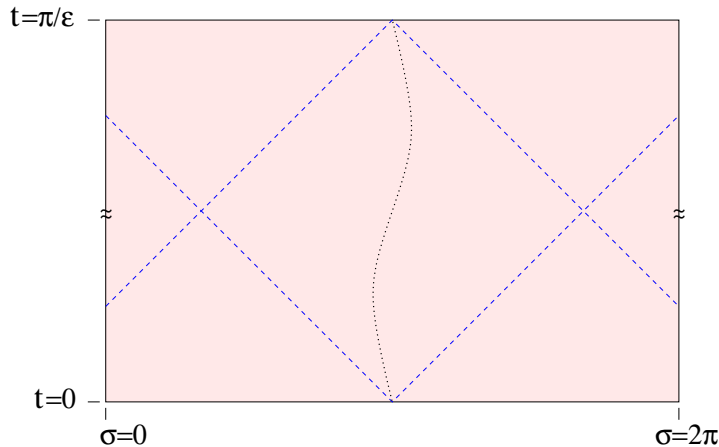}
\end{picture}}
\]
\caption{\it 
Conformal diagram for a spacetime which is asymptotically
de~Sitter at both past and future infinity.
The worldline of a hypothetical observer in the
spacetime is indicated, together with its past
and future horizons.  As $\epsilon$ decreases, the observer
has access to more of the spacetime.
}
\end{center}
\end{figure}

\subsection{\label{liourev}de~Sitter Liouville}

Let us begin with a review of the classical theory, following%
~\cite{DaCunha:2003fm}.  The notational conventions of the
present paper conform to the conventions of%
~\cite{Fateev:2000ik,Teschner:2000md,Zamolodchikov:2001ah},
which are different from those of~\cite{DaCunha:2003fm};
the translation between these two sets of conventions may be found 
in appendix~\ref{conventions}.
Working above the critical dimension involves a continuation
of $b$ to imaginary values; therefore let us define
\be
  \beta=-ib\quad,\qquad 
  \phi=i\varphi\quad.
\label{rotate}
\ee
The general classical solution of the Liouville theory~\pref{sliou}
can be expressed locally as
\be
  e^{2\beta\phi} = -\frac{1}{\pi\mu\beta^2}
        \;\frac{\partial A(x^+)\partial B(x^-)}{[A(x^+)-B(x^-)]^2}\quad ,
\label{gensoln}
\ee
where $x^\pm=t\pm\sigma$.  The gauge constraints
\be
  T_{\pm\pm}^{\rm Liouville} + T_{\pm\pm}^{\rm matter} = 0
\label{vircon}
\ee
are a set of non-linear differential equations that
determine $A(x^+)$ and $B(x^-)$
in terms of the matter stress-energy.
One may broadly characterize the solutions according
to the monodromy of the functions $A$ and $B$
\be
  A(x^+\!+2\pi) = \frac{aA(x^+)+b}{cA(x^+)+d}\quad,\qquad
  B(x^-\!-2\pi) = \frac{aB(x^-)+b}{cB(x^-)+d}
  \quad ;
\label{monodromy}
\ee
the solution is said to be of the elliptic, parabolic, or hyperbolic
class, depending on whether the trace of the monodromy matrix
${\bf M}=({a~b\atop c~d})$ has 
${\it Tr}\,{\bf M}<2$, ${\it Tr}\,{\bf M}=2$, or ${\it Tr}\,{\bf M}>2$, 
respectively.  In string theory terms, these solutions
correspond to tachyonic, massless,
and massive strings in the target space, respectively;
they also correspond to positive, zero, or negative
Liouville energy, see figure~\Lpot.

\begin{figure}[ht]
\begin{center}
\[
\mbox{\begin{picture}(318,136)(0,0)
\includegraphics{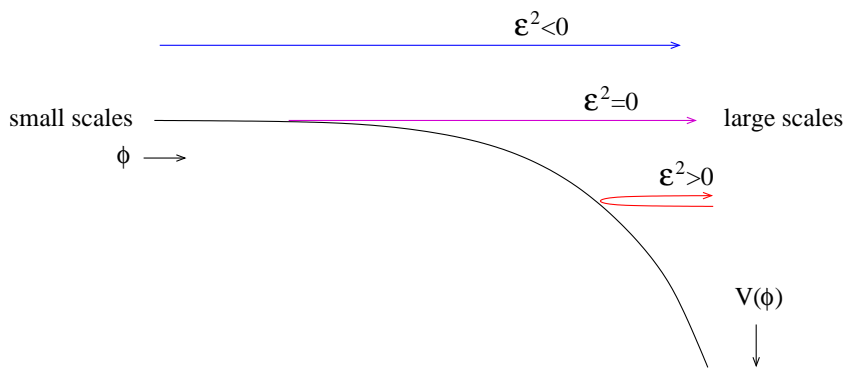}
\end{picture}}
\]
\caption{\it The three classes of solution of de~Sitter Liouville theory,
corresponding to negative, zero, and positive energy
$E_L+\frac{Q^2}{8}=\frac{\epsilon^2}{2\beta^2}$
relative to the Casimir shift.
The Liouville field has `wrong sign' kinetic energy
(because it is the conformal factor of the metric);
the sign of the potential has been flipped here for the
purpose of using standard intuition about potential dynamics.
}
\end{center}
\end{figure}

The elliptic category of classical solutions to 
de~Sitter Liouville theory are asymptotically de~Sitter
in both past and future; they reach infinite scale factor
at finite conformal time parameter in both past and future.
An example of such a space is the metric
\be
  ds^2 = e^{2\beta\phi}(-dt^2+d\sigma^2)
	= \frac{1}{\pi\mu \beta^2}\,\frac{\epsilon^2}{\sin^2(\epsilon t)}
	\;(-dt^2+d\sigma^2)\quad ,
\label{globaldsds}
\ee
where $t\in(0,\pi/\epsilon)$ and $\sigma\in(0,2\pi)$;
the conformal diagram of this geometry is shown in figure~\dspm.
In the parametrization~\pref{gensoln}, one has
$A(x^+)=\tan(\hf\epsilon x^+)$ and $B(x^-)=-\tan(\hf\epsilon x^-)$.
The Hamiltonian (Virasoro, BRST) constraints fix the 
parameter $\epsilon$ in terms of the matter energy
\be
  \frac{\epsilon^2}{2\beta^2}+E_{\rm matter}-\frac{d-1}{12}=0 \ .
\label{lzero}
\ee
Global de~Sitter space corresponds to $\epsilon=1$;
increasing $E_{\rm matter}$ decreases $\epsilon$ until
one reaches the threshold of cosmological singularity formation
at $\epsilon=0$, where the Liouville solution is then
of the parabolic class.  A slight further increase in $E_{\rm matter}$
changes the global geometry; $\epsilon = i\varepsilon$
becomes purely imaginary.  The Liouville solution is
then of the hyperbolic class, and exhibits a Milne-type 
cosmological singularity.
For more details on the classical solutions, see%
~\cite{DaCunha:2003fm}.


The fact that $\phi\to\infty$ on the asymptotically de~Sitter
conformal boundary of the classical spacetime suggests that the appropriate
quantum analogue uses the boundary states~\pref{mnvac}.
Roughly, the boundary states~\pref{mnvac} in this regime of central charge
$c$ describe S-branes located in the vicinity of $\phi=\infty$.
The fact that the conformal diagram of the classical spacetime
is a finite cylinder suggests that we consider
the annulus amplitude with these boundary states.

\subsection{\label{dsannamps}Liouville theory as a 2d dS CFT}

Consider the Liouville annulus partition function~\pref{lann}
coupled to $c=d$ matter and ghosts; for example, one might take
the matter to be $d$ free scalar fields.  The combined
partition function is
\bbb
  \ZZ(\pp,mn|-\!\pp,m'n') &=& 
\label{dsann}\\
  & &\hskip -4.5cm
\int_0^\infty\!\!d\tau\!\!\int_{-\infty}^\infty\!\!\!\!\!{d\nu}\, 
	\frac{\qq^{\frac14\pp^2+\nu^2}}{[\sqrt2\,\eta(\qq)]^{d-1}}\;
	\frac{2\,\sin(2\pi m\nu/\beta)\sin(2\pi n\nu \beta)\cdot
		2\,\sin(2\pi m'\nu/\beta)\sin(2\pi n'\nu \beta)}
	{2\sin(2\pi\nu/\beta)\sin(2\pi\nu \beta)}\ ,
\nonumber
\eee
with $\nu = \epsilon/2\beta$.  
More generally, the matter plus ghost contribution will have the structure
\be
  \frac{\qq^{\frac14\pp^2}}{\sqrt 2\,[\sqrt 2\,\eta(\qq)]^{d-1}}
	~\longrightarrow~ 
	\int_0^\infty\! d\omega \,\rho(\omega)\,
	 \qq^{\omega-\frac{c-1}{24}}\ .
\label{genmatter}
\ee
For simplicity, consider $m'=n'=1$.  
The amplitude becomes
\be
  \ZZ(\pp;m,n) = 
	\int_0^\infty\!\!\!\!d\tau
	\int_{-\infty}^\infty\!\!\!\!\!{d\nu}
	\int_0^\infty\!\!\!\!\!d\omega\,\rho(\omega)\,
        \qq^{\frac14\pp^2+\nu^2+\omega-\frac{c-1}{24}}\cdot
	2\sqrt2\,\sin(2\pi m\nu/\beta)\sin(2\pi n\nu \beta)
\label{mnone}
\ee
The integral over $\nu$ is Gaussian; 
note that it has saddle points located at 
\be
  \nu = \pm\frac{i(m/\beta\pm n\beta)}{2\tau}\ ,
\label{saddle}
\ee
and giving the result
\bbb
  \ZZ(\pp;m,n) &=&
        \int_0^\infty\!\!\!\!d\tau 
        \int_0^\infty\!\!\!\!\!d\omega\,\rho(\omega)\,
        \qq^{\frac14\pp^2+\omega-\frac{c-1}{24}}\cdot
        2\sqrt2
\label{steptwo}\\
	&&\hskip 1cm
	\times\Bigl(
		\exp[-\frac{\pi}{2\tau}(m/\beta-n\beta)^2]
		-\exp[-\frac{\pi}{2\tau}(m/\beta+n\beta)^2]
	\Bigr)
\nonumber
\eee
The integral over $\tau$ is manifestly divergent
for $\frac14\pp^2+\omega<\frac{c-1}{24}$,
for a trivial reason -- we have been attempting to use 
Euclidean worldsheets to describe an intrinsically Lorentzian
process, namely what is from the target space (string) perspective
the propagation of closed string tachyons.
One has improperly attempted to rotate the integral
over the Schwinger parameter from an oscillating exponential
to a real exponential.
Therefore let us return the $\tau$ integral back to the 
imaginary axis from which it originated, defining $\tau=i\tim$.
The resulting integral over $\tim$ is elementary, and yields
\bbb
  \ZZ(\pp,mn) &=& 
	\int_0^\infty\!\!\!\!\!d\omega\,\rho(\omega)\,
	\Bigl(\frac{{i}/{\sqrt2}}{\;
	[\frac{c-1}{24}-\frac14\pp^2-\omega]^{1/2}}\Bigr)\;
\label{dsprop}\\
	&&\hskip 1cm
	\times\Bigl[\exp\Bigl(2\pi i(m/\beta-n\beta)
		[\coeff{c-1}{24}-\coeff14\pp^2-\omega]^{\half}\Bigr)
		-(n\to -n)\Bigr]\ .
\nonumber
\eee
Note that the dominant contribution to the $\tim$
integral comes from the saddle point
\be
  \tim = \frac{(m/\beta\pm n\beta)}%
	{2\,[\coeff{c-1}{24}-\coeff14\pp^2-\omega]^\half}\ ,
\label{bitsa}
\ee
or in other words
\be
  \nu^2=\coeff{c-1}{24}-\coeff14\pp^2-\omega
\label{mashell}
\ee
which is the mass shell condition~\pref{lzero}
for the propagating string mode.  Plugging into~\pref{saddle},
we see that the classical geometry~\pref{globaldsds}
is recovered in the limit $\beta\to 0$, provided 
once again that $m=1$.  It may be that further
consideration of the loop perturbation expansion in 
Liouville theory around the classical solution~\pref{globaldsds}
might only match the exact answer~\pref{dsprop}
for $m\!=\!n\!=\!1$, as in~\cite{Fateev:2000ik}.

In the semi-classical limit $\beta\to0$, one expects
that fluctuations around the classical geometry are
suppressed, at least for smooth spacetimes.  Consider
a double saddle point approximation to the $\nu$ and $\tim$
integrals; as mentioned above, the saddle points are located
at~\pref{saddle} and~\pref{bitsa}.  The width
of these saddles is (dropping subleading terms in $\beta$)
\be
  \frac{\delta\tau}{\tau} \sim \frac{\delta\nu}{\nu}
	\sim \frac{\beta}{[\pi^2m^2(1-(\pp^2+4\omega)\beta^2)]^{1/4}}\quad.
\label{width}
\ee
Thus, as long as we are far from the threshold of the
cosmological singularity at $\nu=0$
(\ie\ $\pp^2+4\omega=\beta^{-2}$), fluctuations in
the geometry are indeed suppressed; we recover
the classical de~Sitter geometry~\pref{globaldsds},~\pref{lzero}
via the strongly peaked saddle point~\pref{saddle},~\pref{mashell}.  
Near the threshold at $\nu=0$,
there are large fluctuations in the path integral
and the classical geometry cannot be trusted.%
\footnote{Note, however, that well above the big bang/crunch
threshold, fluctuations are again suppressed,
but the saddle point is off in the complex plane.
This fact will be important below when we discuss factorization
of the annulus.}

\subsection{\label{dswavefn}Wavefunctions}

We now turn to a discussion of de~Sitter wavefunctions.
The transform~\pref{backlund} of the $c\le 1$ Liouville
wavefunctions~\pref{Lfunction} in $\sigma$ space yields the
wavefunctions for the boundary length $\ell=\oint e^{b\varphi}$,
namely the Bessel functions $K_{iE}(M\ell)$.
The analogue for de~Sitter Liouville should reproduce the
corresponding de~Sitter wavefunctions, which in the
minisuperspace approximation can be taken to be
Hankel functions (see \eg~\cite{DaCunha:2003fm}).
The analogue of the transform~\pref{backlund} for Hankel functions is
\bbb
  \coeff{i\pi}{2}\,e^{-\pi E/2}H_{iE}^{\sst(1)}(M\ell) 
	&=& \int_0^\infty \!du\; e^{iM\ell\,\cosh \,u}
		\,\cos(Eu)
\nonumber\\
  \frac{-2i\,e^{\pi E/2}}{E\,\sinh(\pi E)}\;\cos(Eu)
	&=& \int_0^\infty\!\frac{d\ell}{\ell}\,
		e^{iM\ell\,\cosh\, u}\,
		H_{iE}^{\sst(1)}(M \ell)\quad .
\label{Hbacklund}
\eee
Indeed, in the continuation~\pref{rotate}, the boundary interaction
which is the kernel of the integral transform~\pref{backlund}
continues as
\be
  \exp[-\mub\ell] ~\longrightarrow~
	\exp\Bigl[\pm i\,\Bigl(
		\frac{\sqrt\mu\,\cosh(\pi\beta s)}{\sqrt{\sin(\pi\beta^2)}}
		\Bigr)\,\ell\Bigr]\quad ,
\label{bdycont}
\ee
where we have also chosen to continue $s=i\sigma$;
the wavefunction~\pref{Lfunction} rotates as
\be
  \Psi_{i\nu}(\sigma) ~\longrightarrow~
	\frac{\Gamma(1-2i\nu \beta)\Gamma(1+2i\nu /\beta)
	\cos(2\pi s \nu )}%
		{2^{1/4}\;(2\pi \nu) }\;\mm^{-i\nu /\beta}\ ,
\label{psicont}
\ee
and similarly for the $(m,n)$ boundary state wavefunctions~\pref{mnvac}.
From~\pref{Hbacklund}, the transform of the wavefunction~\pref{psicont}
to $\ell$ space can be written
\be
  \Psi_{i\nu}(\ell) 
	= \left[\coeff{i\Gamma(1\!-\!2i\nu\beta)}{2^{1/4}\,2\pi\,\beta}\,
	\Bigl(\coeff{\Gamma(1\!+\!2i\nu/\beta)}%
		{\Gamma(1\!-\!2i\nu/\beta)}\Bigr)^{\!\half}\right]
		\;(-\mm)^{-i\nu/\beta} 
	\sqrt{\pi(2\nu/\beta)\,\sinh(2\pi\nu/\beta)}\;
		H_{2i\nu/\beta}^{\sst(1)}(M\ell)
\label{dsellfn}
\ee
with $M=\pi\sqrt{\mu/\sin(\pi\beta^2)}=\Gamma(1\!+\!\beta^2)\sqrt{-\mm}$.
These are indeed, up to normalization,
the appropriate minisuperspace de~Sitter wavefunctions
in the semi-classical limit $\beta\to0$.
These wavefunctions arise in the disk one-point functions
of vertex operators with hyperbolic monodromy,
which are the continuation of equation~\pref{onepoint}.
Thus the appropriately normalized disk one-point functions
provide the transition amplitudes for the hyperbolic class
of geometries that are only asympototically de~Sitter in
the future (or past), and have a big bang singularity
in the past (or future).  The vertex operator at the puncture in the disk
specifies the state of the universe at the big bang (or crunch).

\subsection{\label{remarks}Comments}

At this point a number of remarks about the result~\pref{dsprop}
are in order.  
\begin{itemize}
\item
The integral over $\omega$ is divergent.
Consider $m\!=\!n\!=\!1$; then the first exponential in
square brackets is
\be
  \exp\Bigl[-2\pi\sqrt{\Bigl(\frac{c-25}{6}\Bigr)
	\Bigl(L_0-\frac{c-1}{24}\Bigr)}\Bigr]\ .
\label{suppression}
\ee
In the semi-classical limit,
this is exactly the suppression factor for the production of
closed strings found in~\cite{DaCunha:2003fm,Strominger:2003fn};
it was also found there that this suppression factor was
overwhelmed by the density of states $\rho(\omega)$,
leading to a divergence in the production of string modes.
Now, standard D-brane boundary states such as
the free field Dirichlet/Neumann states
\be
  \ket{B} = \exp\Bigl[\pm\sum_{n>0} 
	\frac1n\,\alpha_{-n}\tilde\alpha_{-n}\Bigr]\ket0
\label{mattbdy}
\ee
do not couple to all closed string states, but
rather only to those with a matching spectrum of
left- and right-moving modes.  These have only
an open string degeneracy
\be
  \rho(\omega)\sim\exp\Bigl[+2\pi\sqrt{\Bigl(\frac{c-1}{6}\Bigr)
        \Bigl(L_0-\frac{c-1}{24}\Bigr)}\Bigr]\ ,
\label{opengrowth}
\ee
but this is still slightly overcomes the suppression factor~\pref{suppression}.
However, we are not interested in actually carrying out the
sum over states; rather, we regard the propagator~\pref{dsprop} 
as a generating function for de~Sitter spacetimes.
We can select the initial and final states of interest
by appropriate change in the choice of matter wavefunctions
on the two boundaries.  We have used D-brane wavefunctions, 
which overlap with a substantial number of (closed string) matter states.
To pick out a particular (closed string)
state of the matter from this set, one should use the
corresponding wavefunction of that state rather than
the D-brane boundary state.

\item
The majority of matter states
do not belong to the partition sum~\pref{dsann}; absent are the
matter states whose left- and right-moving excitations are different.  
For a general matter state, it seems appropriate to use
the Ishibashi state associated to the Virasoro module
of a matter highest weight state.  The matter contribution
to the annulus amplitude is then a Virasoro character
such as~\pref{liouchar}, and as in section~\ref{minmod},
longitudinal modes cancel among Liouville, matter, and ghost sectors.
The annulus amplitude will then take the form
of the integrand in~\pref{dsprop}.

\item
The equivalence of the exponential factor in~\pref{dsprop}
with the pair production probability amplitude~\pref{suppression}
is not an accident.  The production probability amplitude
for massive string modes is essentially the above-barrier reflection
amplitude in the Liouville potential of figure~\Lpot,
see for instance~\cite{Audretsch:1979uv}.
We see that the annulus amplitude~\pref{dsann}
computes the probability that the universe starts
off at infinite scale factor and returns to infinite scale factor,
even if the energy is such that the geometry classically
wants to climb onto the flat part of the Liouville potential
and shrink to zero scale factor (the time reverse of the 
negative energy solution of figure~\Lpot).

\item
An alternative, standard way to obtain general amplitudes 
uses perturbations of the state 
at the conformal boundary~\cite{Gubser:1998bc,Witten:1998qj}
(for a review, see~\cite{Aharony:1999ti}; for a discussion
in the de~Sitter context, see~\cite{Witten:2001kn,Strominger:2001pn}).  
In fact, the AdS version of our discussion involves exactly this
prescription.  Matter perturbations build up an arbitrary state
of the matter CFT, which is then dressed 
by the Liouville gravity/ghost sector.
The $m\!=\!n\!=\!1$ Liouville state has only the Liouville identity
operator (and its Virasoro descendants) at the boundary,%
\footnote{For instance, taking a Liouville closed string vertex
operator to the boundary yields only the identity operator 
and its descendants as the vertex interacts with its image
across the boundary, \cf~\cite{Zamolodchikov:2001ah}.} 
\ie\ the state is always asymptotically locally (anti)de~Sitter.

For the general $(m,n)$ vacua with $m,n\ne 1$,
there is the further possibility of adding 
boundary vertex operators $B_i$ on either end of the annulus
\be
  \Bigl\langle{\prod_{i=1}^{N_{\it in}}
	\oint_{\partial_{\it (in)}}\!\!\!\!\! B_i\;
	\prod_{j=1}^{N_{\it out}}
	\oint_{\partial_{\it (out)}}\!\!\!\!\!\! B_j}
	\Bigr\rangle\ .
\label{transamp}
\ee
For example, $B=e^{\alpha\phi+i\pp\cdot\xx}\,\partial^k\!X$ 
adds or subtracts an excitation of the $k^{\rm th}$ mode
of the free matter field $X$.  Similarly, we can perturb
the disk amplitudes that give the transition amplitudes
for spaces that are asymptotically de~Sitter in only the
past or future.
Now, the boundary condition
associated to the $(m,n)$ boundary state admits
nontrivial correlations with only a finite number 
of Liouville boundary primaries, see equation~\pref{fusion}.
The Liouville exponent $\alpha$ of matter boundary vertex operators
is generically not compatible with the null vector condition
that leads to $\sigma=\sigma(m,n)$ in~\pref{bsrel},
\ie\ $\alpha\not\in\{\alpha_{m,n}=(1\!-\!n)\beta-(1\!-\!m)/\beta\}$.
Of course, one could always adjust the matter momentum $\pp$
in the free field case in order to satisfy the constraints
with $\alpha=\alpha_{m,n}$ for some $m$, $n$,
but this seems rather restrictive.  In any event,
there are only a rather small number of perturbations
by integrated boundary vertex operators
of the general $(m,n)$ vacua, and boundary perturbations
of the standard global vacuum state with $m\!=\!n\!=\!1$
cannot involve the Liouville field.

\item
Recently, the mere existence of de~Sitter quantum gravity 
has come into question%
~\cite{Dyson:2002nt,Dyson:2002pf,Goheer:2002vf}.
The complaints come in several versions: 
\begin{itemize}
\item
(1) Correlation functions cannot decay completely to zero,
contradicting locality;
\item
(2) Poincar\'e recurrences lead to bizarre and unacceptable cosmology;
\item
(3) Finiteness of the density of states is incompatible with 
the classical $SO(d,1)$ isometry of the $dS_d$ vacuum.
\end{itemize}
The first two of these do not preclude the existence of
de~Sitter quantum gravity; they are more aesthetic in nature.  
The last assertion seems more serious.  
The isometry in question for $dS_2$ is $SO(2,1)\simeq SL(2,R)$.
This $SL(2,\IR)$ is the symmetry
of the global $\epsilon=1$ vacuum solution~\pref{globaldsds}.%
\footnote{Another $SL(2,\IR)$ is the
monodromy~\pref{monodromy} of the classical solutions.
In the quantum theory, this becomes a
hidden, internal quantum group symmetry
$\UU_q({{\mathfrak s}{\mathfrak l}}(2,\IR))$%
~\cite{Gervais:1990di,Cremmer:1997eh,Ponsot:1999uf,Ponsot:2000mt}
(for reviews, see~\cite{Gervais:1992td,Teschner:2001rv,Ponsot:2003ju}).
Liouville chiral vertex operators lie in well-defined representations
of this symmetry, which is thus maintained 
(although $q$-deformed) in the quantum theory.} 
In the quantum theory, this becomes the $SL(2,\IR)$ symmetry
of the $m\!=\!n\!=\!1$ Liouville open string vacuum state
(which formally has a null vector at level one,
\ie\ is annihilated by $L_{-1}$ in addition to $L_1$ and $L_0$);  
thus this symmetry is preserved in the quantum theory.
It would be interesting to understand the relation between
the arguments of%
~\cite{Dyson:2002nt,Dyson:2002pf,Goheer:2002vf}
and the construction of quantum de~Sitter space in two dimensions
that we have presented here.
One difference of the present analysis with that of%
~\cite{Dyson:2002nt,Dyson:2002pf,Goheer:2002vf}
is the assumption in those works of a well-defined Hilbert space and quantum
mechanics for the `static patch' of de~Sitter space;
the constructions of the present article uniformly employ a global
description.

\item
We have implicitly assumed that the quantum matter states
of interest are built on the conformal vacuum of
the matter CFT.  Recently, there has been some interest
in considering other classes of de~Sitter invariant matter states%
~\cite{Mottola:1985ar,Allen:1985ux}, in particular in
connection with possible `trans-Planckian' effects in inflationary
cosmology.  These states $\ket\alpha$ are defined by the conditions
\bbb
  (\alpha_n\,\cosh\,\alpha-\alpha_{-n}\,\sinh\,\alpha)\ket\alpha &=& 0
\nonumber\\
  (\tilde\alpha_n\,\cosh\,\alpha
	-\tilde\alpha_{-n}\,\sinh\,\alpha)\ket\alpha &=& 0\ ,
\label{alphavac}
\eee
or in other words
\be
  \ket\alpha = \exp\Bigl[\hf\,\tanh\,\alpha\,
	\sum_{n>0}\frac1n(\alpha_{-n}^2+\tilde\alpha_{-n}^2)\Bigr]\ket 0
\label{alphastate}
\ee
for the oscillator modes of a free field
(compare~\pref{mattbdy}).
It seems problematic to consistently couple such states, 
which (like D-brane boundary states)
lie outside of the CFT Hilbert space, to Liouville gravity.  
In particular, they do not seem to admit a sensible action
of the conformal algebra,
so it seems difficult to satisfy the BRST constraints.
In any event, two dimensional de~Sitter
gravity seems a fruitful laboratory for the search for novel
properties of quantum matter and gravity in inflating spacetimes.

\item
The homogeneous classical solution to de~Sitter Liouville theory
in the negative energy, hyperbolic class 
of monodromies~\pref{monodromy} is
\be
  e^{2\beta\phi} = \frac{1}{\pi\mu\beta^2}
	\;\frac{\varepsilon^2}{\sinh^2(\varepsilon t)}\ .
\label{hypsoln}
\ee
If one takes this solution seriously for all times $t\in(-\infty,\infty)$,
the $t<0$ region describes a big bang at $t=-\infty$ that
expands to an asymptotically de~Sitter geometry at $t\to0^-$, 
and $t>0$ describes a big crunch that starts at asymptotically
de~Sitter space at $t\to0^+$ and crunches at $t=+\infty$.
If we formally compactify $t$ by adding the point at infinity,
one might ask whether the resulting crunch/bang spacetime appears
in the annulus amplitude~\pref{dsprop}.  
In other words, does the passage from positive to negative
Liouville energy look like figure~\pinch?

\begin{figure}[ht]
\vskip -.6cm
\begin{center}
\[
\mbox{\begin{picture}(223,141)(0,0)
\includegraphics{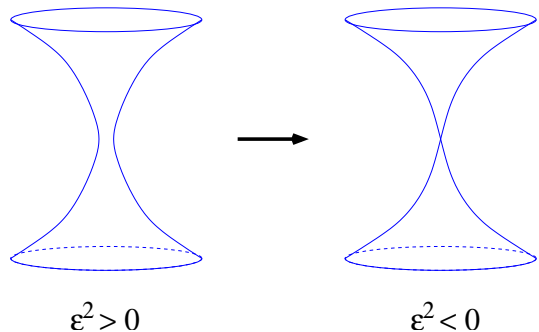}
\end{picture}}
\]
\vskip -.3cm
\caption{\it 
Naively, passage from the elliptic to the hyperbolic
class of Liouville solution leads to a big crunch/big bang geometry.
}
\end{center}
\vskip -.2cm
\end{figure}

The exponential
suppression of the amplitude at negative Liouville energy
(\ie\ matter energy larger than $\frac{c-1}{24}$),
together with its consistent interpretation in terms of
above barrier reflection in the de~Sitter Liouville potential,
seems to indicate that the answer is no.  
Is there any way to obtain the crunch/bang geometry of
figure~\pinch?  Naively, such a geometry should arise if
we pinch the annulus, factorizing on the disk one-point
amplitudes (the unnormalized version of~\pref{onepoint})
for localized vertex operators $e^{2\alpha\phi}$,
$\alpha\in \frac Q2+i\IR$,
that create the negative energy Liouville states of interest.
The fact that the spatial circle shrinks to a point at the
location of the vertex operator is associated
to an infinitely long propagation in conformal time;
the Liouville vertex operator specifies the 
nature of the wavefunction of the geometry in the region of
small scale factor.  
Ordinarily, the pinch of the annulus does correspond
to the limit $\tim\to\infty$, where closed
string propagation goes on shell;
and the disk one-point amplitudes are the residue
of the closed string propagator pole.
However, this conventional intuition actually breaks down here.
To see the source of the problem, let us perform the $\nu$
and $\tim$ integrals in the opposite order.  Doing the $\tim$
integral in~\pref{dsann}
first reveals the pole structure in the amplitude
\bbb
  \ZZ(\pp,mn) &=& 
	i\int_{-\infty}^\infty\!\frac{d\nu}{2\pi i}
	\int_0^\infty\!\!\!\!\!d\omega\,\rho(\omega)\,
		\frac{1}{\frac14\pp^2+\nu^2+\omega-\frac{c-1}{24}}\;
\label{taufirst}\\
        &&\hskip .7cm
		\times\;\frac{2\,\sin(2\pi m\nu/\beta)\sin(2\pi n\nu \beta)
		\cdot 2\,\sin(2\pi m'\nu/\beta)\sin(2\pi n'\nu \beta)}%
		{\sqrt2\, \sin(2\pi \nu/\beta)\sin(2\pi \nu \beta)}\quad .
\nonumber
\eee
Now, for $\frac14\pp^2+\omega>\frac{c-1}{24}$, we might 
try to close the contour at infinity to pick out the 
residue of the pole at $\nu=i\hat\nu$, where
$\hat\nu=(\frac14\pp^2+\omega-\frac{c-1}{24})^{1/2}$, to get
\be
  \frac{2\,\sinh(2\pi m\hat\nu/\beta)
		\sinh(2\pi n\hat\nu \beta)
  \cdot 2\,\sinh(2\pi m'\hat\nu/\beta)
		\sinh(2\pi n'\hat\nu \beta)
		}{\sqrt 2\,2\hat\nu
	\,\sinh(2\pi \hat\nu/\beta)
                \sinh(2\pi \hat\nu \beta)}\ .
\label{bogus}
\ee
The product of the disk one-point functions with
$(m,n)$ and $(m',n')$ boundary conditions is indeed of 
this form.  The problem, of course, is that
closing the contour in this way is improper due
to the presence of the sine functions; the correct closing
of the contour always selects the exponentially damped
contribution, as in~\pref{dsprop}, rather than the
sinh that would be required to reproduce the 
factorization on the crunch/bang geometry.
Thus the annulus amplitude can only yield
the above-barrier reflection amplitude, and 
does not tell us how to propagate through
a big crunch geometry and emerge on the other side
into a big bang.

\item
One might ask why and how big bang or big crunch singularities
are regularized in two-dimensional gravity.
It is essentially a result of conformal invariance;
we are specifying the particular UV completion of the 
theory that defines the physics in the neighborhood
of the `singularity', which from the point of view of
two-dimensional CFT is a boundary condition at a puncture
describing the `big bang' state there.  
The partition functions of such geometries are disk amplitudes such as
\be
  \vev{e^{2\alpha\phi}\,\OO_{\rm matter}}^{ }_{(m,n)}
\label{bigbang}
\ee
with asymptotically de~Sitter boundary conditions~\pref{mnvac};
the Liouville part of these amplitudes
is the continuation of~\pref{onepoint},
associated to the wavefunction~\pref{dsellfn}.
This is not a trivial statement;
when one couples $2d$ gravity non-trivially to matter,
as in the inflationary cosmologies studied 
in~\cite{DaCunha:2003fm}, the dynamics changes
dramatically from small scale (large negative Liouville field $\phi$)
to large scale (large positive $\phi$).
At large positive $\phi$, there can be all kinds of complicated
matter interactions, dressed by gravity.
At large negative $\phi$, however, the dynamics must
converge to some particular matter CFT coupled to Liouville,
and in addition all the Liouville dressings to
relevant matter interactions $e^{2\alpha\phi}\OO_{\rm matter}$
shut off (irrelevant matter interactions typically do not lead to
a well-defined continuum theory).
Thus the theory at small scales is 
specified by the choice of a conformal field theory.
Presumably in higher dimensions it is the UV completion
of quantum gravity plus matter, \ie\ string theory,
that plays the same role in taming cosmological singularities.

\end{itemize}

Finally, in the light of the connection between $c_{\rm matt}\le 1$
noncritical string theory and matrix models, it is rather
intriguing that the de~Sitter Liouville gravity dynamics obtained 
in the regime $c_{\rm matt}\ge 25$ is so closely parallel.
It hints at a dual formulation of two-dimensional quantum
de~Sitter gravity in terms of a matrix model.  The asymptotic
de~Sitter boundary condition is implemented by what
could reasonably be described as `S-branes at infinity'
in $\phi$ space.%
\footnote{Which in some sense `live' on the boundary
of spacetime, much as in the AdS/CFT correspondence.}
One might hope that these are the
worldsheet reflection of matrix degrees of freedom,
just as in the $c< 1$ disk instanton amplitudes of
section~\ref{infbranes}; and that careful consideration
of de~Sitter Liouville amplitudes of the sort that we
have constructed here will lead us to the appropriate
matrix theory.


\vskip 2cm
\noindent
{{{\bf Acknowledgments}}}:
Thanks to
D. Kutasov
and 
W. McElgin
for discussions,
and
G. Moore
for a correspondence.
This work was supported by DOE grant DE-FG02-90ER-40560.


\appendix
\section{\label{conventions}Conventions}

The present paper follows the conventions of%
~\cite{Fateev:2000ik,Teschner:2000md,Zamolodchikov:2001ah},
which differ from those of%
~\cite{Ginsparg:1993is,DaCunha:2003fm}.
Let us denote quantities appearing in the latter
works by the subscript {\it GM}.  The present paper,
as well as~\cite{Fateev:2000ik,Teschner:2000md,Zamolodchikov:2001ah},
sets $\alpha'=1$, whereas $\alpha'_{\sst \it GM}=2$.
The parameters of the Liouville field theory are related by
\bbb
  b &=& {\gamma_{\sst \it GM}}/{\sqrt2}
\nonumber\\
  \mu &=& \frac{\mu_{\sst \it GM}}{8\pi\gamma_{\sst \it GM}^2}
\label{liouparams}
\eee



\providecommand{\href}[2]{#2}\begingroup\raggedright\endgroup

\end{document}